\documentstyle[prb,twocolumn,aps,epsf,psfig]{revtex}
\begin{document}
\draft

\twocolumn[\hsize\textwidth\columnwidth\hsize\csname @twocolumnfalse\endcsname

\title{On the soliton width in the incommensurate phase of spin-Peierls
systems.} 

\author{Ariel Dobry and Jos\'{e} A. Riera}
\address{Departamento de F\'{\i}sica, Universidad Nacional de Rosario, \\
and Instituto de F\'{\i}sica Rosario, Avenida Pellegrini 250,
2000 Rosario, Argentina}
\maketitle

\begin{abstract}
We study using bosonization techniques the effects of frustration due 
to competing interactions and of the interchain elastic couplings on
 the soliton
width and soliton structure in spin-Peierls systems. We compare the 
predictions of this study with numerical results obtained by exact 
diagonalization of finite chains.  
We conclude that frustration produces in general a 
reduction of the soliton width while the interchain elastic coupling
increases it.
We discuss these results in connection with recent measurements
of the soliton width in the incommensurate phase of CuGeO$_3$. 
\end{abstract}


\vskip2pc]
\narrowtext

One dimensional (1D) magnetic systems are unstable against a 
spin-lattice interaction. At a finite temperature a phase transition 
takes place toward a dimerized or spin-Peierls (SP) state whose magnetic 
signature is the opening of a gap (SP gap) between the 
singlet ground state and triplet excitations. The recent
 discovery\cite{hase} 
of the first inorganic SP compound CuGeO$_3$ 
has renewed the
interest in these quasi-one-dimensional spin-phonon coupled systems.
A number of experiments can be performed on these inorganic materials
in an easier way than on the organic SP systems. It is
also now possible to replace the Cu ions by other magnetic or
nonmagnetic ions and to study the effects of these 
impurities.\cite{Lussier} Adding interest to the study of this
compound is the fact that, as it was early recognized, 
the temperature dependence of the spin
susceptibility could only be accounted for if an important 
next nearest-neighbor (NNN) is included in the 1D
Heisenberg model used to describe this material.\cite{RieraD,Castilla}
The dispersion of magnetic excitations and a number of other
experimentally determined properties are also consistent with a  
model with NNN interaction.\cite{Haas,RieraK,Poilblanc}

Recent x-ray-scattering experiments have detected a transition from 
the uniform SP phase into an incommensurate
phase\cite{Kiry} in the presence of a magnetic field.
The experimental x-ray spectra in the
incommensurate state has been interpreted as due to a soliton lattice
structure. A soliton deformation was early theoretically predicted 
for SP systems\cite{NF}. It arises because the Zeeman 
energy favors states with 
nonzero magnetic moment thus breaking the dimer pattern. That is,
each down spin changed into an up spin breaks a singlet dimer and
gives rise to two solitons, each soliton carrying a spin-$\frac 12$.

The soliton width has been experimentally estimated from the 
relationship between the measured intensities of the main x-ray peak 
and its harmonics.\cite{Kiry} 
The resulting value is $\xi=13.6 c$ with $c$ the lattice constant in 
the spin chain direction ($c=1$ hereafter).
The predicted width for a one-dimensional model with only nearest 
neighbor (NN) interaction is given by $\xi=J \pi/(2\Delta)$, where
J is the exchange coupling and $\Delta$ the SP 
gap.\cite{NF}
Using the experimental value of $\Delta=2.1\; \rm meV$ and 
$J=120\; \rm K$, estimated by fitting the spin susceptibility
with a Heisenberg model with NN interactions, the
predicted $\xi$ is equal to $8$ which is considerably 
smaller than the experimental one.

It is then quite apparent that some other ingredients should be 
added to the model in order to explain the origin of this 
disagreement between theory and experiment.
One
of the goals of the present work is to analize the effect of the 
frustration induced by the competing NN and NNN antiferromagnetic 
interactions on the
soliton width. In this sense, we are going to extend the 
formalism which leads to the
relationship between the soliton width and the gap, originally
developed in Ref. [\onlinecite{NF}].

Another possible contribution to the soliton width comes from the
interchain elastic coupling. It is well known that 
the three-dimensional (3D) character of the phonon field is essential
to account for the finite temperature SP transition.
Moreover, for the specific case of  CuGeO$_3$, 
it has been found that the 
principal lattice anomaly takes place in a direction perpendicular
to the spin chain\cite{Lorenzo,Hirota} and it is related with 
displacements of the oxygen ions toward and against the chains.
In this situation the phonon modes perpendicular to the chains 
involving coherent motion of the atoms in different chains become
relevant in the SP phase.\cite{Lorenzo} 
It is then reasonable to assume that the
interchain elastic coupling could be relevant to explain some
features of the system. 
The 
possibility that interchain elastic coupling could
modify the soliton shape 
is supported by a recent measurement of the elastic
constants in CuGeO$_3$ in the presence of high magnetic 
fields.\cite{saint-paul} In this study, it was shown that the elastic
constant along the b axis is the one that presents the largest
variation at the dimerized-incommensurate transition.

We conclude that frustration produces in general a 
{\it reduction} of the soliton width while the elastic coupling
of a chain with its neighbors increases it.

We start our study by analyzing the effects of frustration on the 
soliton width.
Therefore, we consider the following 1D model Hamiltonian: 
\begin{equation}
H=H_{el}+H_{mg}
\label{H}
\end{equation}
\begin{eqnarray}
\frac{H_{el}}J=\frac 12K_{\parallel }\sum_i(u_{i+1}-u_i)^2
\nonumber
\end{eqnarray}
\begin{eqnarray}
\frac{H_{mg}}J=\sum_i\left\{ (1+(u_{i+1}-u_i)) {\bf S}_i \cdot 
{\bf S}_{i+1}+\alpha {\bf S}_i \cdot {\bf S}_{i+2}\right\}  
\nonumber
\end{eqnarray}
where $\bf S_i$ are spin-1/2
operators and $u_i$ is the displacement of magnetic ion $i$ from
its equilibrium position. Notice that in CuGeO$_3$, the
displacements $u_i$ are related to the displacements of the O ions.
$H_{el}$ in Eq. (\ref{H}) corresponds to 
the elastic energy interaction of the ions along the
chain with dimensionless elastic constant $K_{\parallel }$.
We include the phonons in the adiabatic 
approximation and we allow arbitrary displacement patterns.

We now go to the bosonic representation of Hamiltonian (\ref{H}). 
This is achieved by first applying a Jordan-Wigner transformation on
the spin operators and then linearizing the resulting fermionic 
relation dispersion around the Fermi level. Finally, the fermionic 
field theory is bosonized. Very precise predictions have been 
obtained within this approach for the ground state and low energy 
excitations of one-dimensional spin systems\cite{Affleck}. 
Nakano and Fukuyama\cite{NF} used this bosonic representation
to study soliton formation in the unfrustrated chain. In the 
following, we extend their approach to the frustrated chain. 
In terms of the boson
field $\phi $ and its conjugate momentum $\Pi $ the Hamiltonian 
(\ref{H}) can be rewritten as:\cite{Zang}

\begin{eqnarray}
H_{el}=2K_{\parallel }\int dxu^2
\label{Hcont}
\end{eqnarray}
\begin{eqnarray}
H_{bos}=\int \frac{dx}{2\pi }\{ \frac{v_s}\eta (\partial _x\phi
)^2+v_s\eta (\pi \Pi )^2+u\sin \phi \nonumber   \\
-g\cos (2\phi )\}
\label{Hbos}
\end{eqnarray}
In going from (\ref{H}) to (\ref{Hcont},\ref{Hbos})
we have replaced the lattice displacement $u_i$ 
by a continuum field as:
\begin{equation}
(-1)^iu_i \rightarrow u(x)
\label{condis}
\end{equation}
so that we can describe smooth deviations of the displacement field
with respect to the dimerized pattern. 

The quantities $v_s$, $\eta $ and $g$ are expressed in terms 
of the microscopic parameters of Hamiltonian (\ref{H}). 
The expressions can be read in Ref. [\onlinecite{Zang}]. 
However, once identified the physical meaning of these 
quantities it is customary to replace those expressions by more precise
relations.\cite{Affleck}
In this sense, the isotropy of the long distance behavior of the 
correlation functions implies that $\eta = 2$.
The parameter $g$ measures the strength of the umklapp processes
in the fermionic theory. The last term 
of (\ref{Hbos}) is a marginally irrelevant operator for all $g > 0$,
so that in absence of coupling to the lattice
a gapless spin liquid phase is predicted.
When g is negative the last term of Eq. (\ref{Hbos}) becomes a 
relevant operator leading the system to a spontaneously dimerized 
gapped phase.\cite{Affleck}
Therefore, $g$ should be proportional to $\alpha_c-\alpha$ where 
$\alpha_c$ is the critical value separating the gapless and
the gapped phases.
The quantity $v_s$ is the spin wave velocity.
For $\alpha=0$, $v_s=J \pi/2$ from Bethe's exact solution. 
For $\alpha < \alpha_c$, $v_s$ has been recently numerically
evaluated as the derivative of the magnon dispersion relation
at the bottom of the band.\cite{Fledder} This derivative is not
well defined for $\alpha > \alpha_c$ due to the presence of the
gap above discussed.
We will analyze later on our procedure to evaluate 
$v_s$ in this case.

The application of a magnetic field to the system favors states with
nonzero $S_z$ and to the appearance of solitons. Each soliton
could be regarded as a domain-wall 
separating two opposite dimerization patterns. To take into account 
these states we follow the approach of Ref. [\onlinecite{NF}]. 
We separate the 
variable $\phi $ into the sum of a classical variable $\phi_0$
and its quantum fluctuation $\hat{\phi}$.
A self-consistent harmonic approximation is then used. We retain 
$\hat{\phi}$ to quadratic order and we require the vanishing of the 
first order term. The displacement field $u$ is obtained from its 
static equilibrium equation. The resulting equations are:

\begin{eqnarray}
-\frac{2v_s}\eta (\partial _x^2\phi )+u\ e^{-\langle \hat{\phi}^2\rangle
/2}\cos \phi _0-2ge^{-2\langle \hat{\phi}^2\rangle }\sin (2\phi _0)=0
\nonumber
\end{eqnarray}
\begin{equation}
\sin \phi_0 e^{-\langle \hat{\phi}^2\rangle /2}+4K_{\parallel }u=0  
\label{ad1d}
\end{equation}
where $\langle \hat{\phi}^2 \rangle $ is the ground state expectation
value which we assume the same as in the dimerized phase. 
These equations could be combined to give: 
\begin{equation}
(\partial _x^2\phi _0)+\frac 1{2\xi ^2}\sin (2\phi _0)=0  \label{soleq}
\end{equation}
The soliton solution of this equation is: 
\begin{equation}
\sin (\phi _0)= \tanh (x/\xi )  
\label{solsol}
\end{equation}
The lattice soliton is then given by:
\begin{equation}
u(x)= u_0\tanh (x/\xi )  
\label{sollat}
\end{equation}
with $u_0$ the equilibrium displacement in the uniform dimerized
state.
The main quantity we are interested in is the soliton width which is 
given by:

\begin{equation}
\frac 1 {\xi^2}
 =\frac \eta {2v_s}\left[ \frac{e^{-<\hat{\phi}^2>/2}}{4K_{\parallel }}
+4ge^{-2<\hat{\phi}^2>}\right].  
\label{solwid}
\end{equation}
This quantity is related to the singlet-triplet gap in the uniform
dimerized state in a simple way. In this state $u(x)=u_0$ 
and $\phi _0=\pi /2$.
By following similar steps as in Ref. [\onlinecite{NF}] we obtain 
the dispersion
relation for the excitations of the system given by:

\begin{equation}
\varepsilon(k)=v_s(k^2+k_0^2)^{1/2}
\label{disrel}
\end{equation}
where $k_0=1/\xi $ is given by Eq. (\ref{solwid}).
Note that $v_s$ is a function of $\alpha$ in this case.
Finally, the simple
scaling relation between the soliton width and the SP
gap $\Delta $,

\begin{equation}
\xi =v_s/\Delta  
\label{width-gap}
\end{equation}
originally obtained for the unfrustrated chain, is also 
valid in the presence of frustration. For
$\alpha > \alpha_c$, $k_0$ contains a contribution from the 
frustration
due to the presence of a gap even in the absence of dimerization.

The relation (\ref{width-gap}) can be tested by numerical exact
diagonalization on finite chains. The adiabatic equations 
corresponding to Hamiltonian (\ref{H}) are:

\begin{equation}
K_{\parallel }\delta_i+\langle 0|{\bf S}_i \cdot 
{\bf S}_{i+1}|0\rangle =0
\end{equation}
where $\delta_i=(u_{i+1}-u_i)$ is the bond length variable. We have 
solved iteratively these equations for $\delta_i$ starting from a 
random configuration $\{\delta_i^{(0)}\}$. 
The ground state of the spin system $|0\rangle $ was
recalculated at each iteration step by Lanczos diagonalization. We 
considered finite chains of even number of sites up to 20 sites with 
periodic boundary conditions.
The numerical details of the method are given
elsewhere \cite{CDFR}.

In the subspace $S^z=0$ the system converges to a dimerized
configuration $\{\delta_i^{(0)}\}$ as expected.
For $\alpha = 0.0$, 0.2 and 0.4, and for several
values of $K_{\parallel }$, using the dimerized pattern 
obtained at $S^z=0$, we computed the spin gap as
$\Delta = E_0(S^z=1) - E_0(S^z=0)$.

When the spin Hamiltonian is diagonalized in the subspace $S^z=1$ we
found that two solitons appear in the dimer pattern. 
The deformation pattern $\{\delta_i\}$ could be described by a 
superposition of two solitons centered at
sites $i_1$ and $i_2$ of the form:

\begin{equation}
\delta_i=(-1)^i\delta_0\tanh \left(\frac{i-i_1}{\xi}\right)
\tanh \left(\frac{i-i_2}{\xi}\right)
\label{fittanh}
\end{equation}
where $\delta_0$, the deformation amplitude,
and $\xi $, the soliton width, were obtained
by numerical fitting. 
The main limitation of this calculation arises in the region where,
for a given $\alpha$, $K_{\parallel }$ is so large that
the solitons have a substantial overlap in our small chain and 
the fitting function (\ref{fittanh}) is no
longer appropriate. Unfortunately, 
it is precisely in this situation where
the continuum approximation leading to the relation (\ref{width-gap}) is 
expected to hold.

\begin{figure}
\psfig{figure=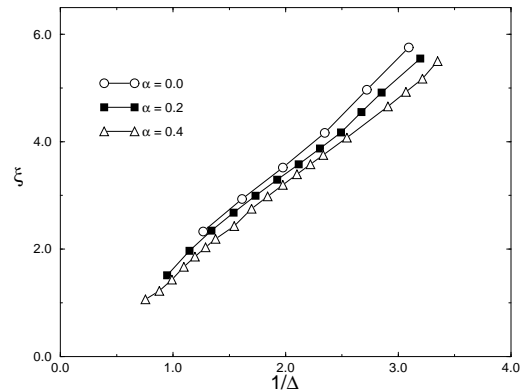,width=8.0cm,angle=-90}
\vspace{0.1cm}
\caption{Soliton width vs. $\Delta^{-1}$ obtained by exact 
diagonalization in the 20 site chain for various values of
$\alpha = J_2/J$.}
\label{fig1}
\end{figure}

In Fig. 1
we show the soliton width $\xi$ as a function of 1/$\Delta $ for
the same values of $\alpha$ as above. The linear
behavior predicted by Eq. (\ref{width-gap}) is clearly seen. 
A linear fitting of these curves in the region $\xi > 2.5$ gives the
slopes 1.87, 1.70 and 1.63 , for 
$\alpha =0.0,\;0.2$ and $0.4$ respectively.
For $\alpha < \alpha_c$, a numerical calculation\cite{Fledder} leads to 
$v_s=\frac \pi 2 (1-1.12 \alpha )$ in the thermodynamic limit. From
this law, $v_s =$ 1.57, 1.22, for 
$\alpha =0.0$ and $0.2$ respectively. We can observe that the slopes
obtained by fitting the numerical data are larger than the calculated
$v_s$. Besides, the effect of $\alpha$ is {\it weaker} in the numerical
data than that predicted by Eq. (\ref{width-gap}). It is difficult
to conclude at this point if this disagreement between the prediction
obtained by the continuum bosonized theory and numerical results
is due to the approximations involved in the former or to finite
size effects.
For $\alpha= 0.4 > \alpha_c \approx 0.2411$, we have estimated $v_s$
by fitting the energy 
$\varepsilon(k)^2 = \Delta^2 + v_s^2 k^2 + c k^4$ around $k=0$,
where $\varepsilon(k) = E_0(S^z=1,k) - E_0(S^z=0,k=0)$. 
For $L=20$, we obtained $v_s = 0.707$, a value which is smaller than
the slope of the curve $\xi$ vs. $1/\Delta$ for $\alpha=0.4$ in
Fig. (\ref{fig1}).

Then the qualitative picture of the effect of frustration is partially
confirmed by
the exact diagonalization study. Both analytically and numerically 
we have seen that frustration reduces the soliton width 
for a given value of the SP gap. However, the numerical calculations
show a smaller dependence on $\alpha$ than that analytically predicted.

\vspace{-0.5cm}

We now turn to the study of the effects caused by the 3D character of the 
phonon field. As previously discussed, there are both theoretical as well
experimental arguments indicating that the interchain elastic coupling 
is important to describe the excitations of the system in the 
SP regime. 
Therefore, we add to $H_{el}$ of Eq. (\ref{H}) an interchain 
elastic coupling of the form: 
\begin{eqnarray}
H_{el}^{\prime }=\sum_{i,{\bf j}} &\{& \frac{K_{\perp }^a}2%
(u_i^{j_x+1,j_y}-u_i^{{\bf j}})^2  \nonumber   \\
&+& \frac{K_{\perp }^b}2
(u_i^{j_x,j_y+1}-u_i^{{\bf j}})^2 \} 
\label{helprime}
\end{eqnarray}
where $u_i^{{\bf j}}$ denotes the displacement of the {\bf j} ion in the 
chain ${\bf j}$ (${\bf j}\equiv (j_x,j_y)$). $K_{\perp }^a$ and 
$K_{\perp }^b$ are
the harmonic couplings in the $a$ and $b$ directions 
respectively. We assume that there is no interchain magnetic
interactions.  The system now consists of a set of 
spin chains immersed in a 3D lattice.
We resort again to bosonization methods to
analyze soliton formation in this system. 
The boson field $\phi$ turns to be ${\bf j}$-dependent. 
The adiabatic condition (\ref{ad1d}) is now given by: 
\begin{eqnarray}
&\sin& \phi _0^{{\bf j}} e^{-<\hat{\phi}^2>/2}+[2(K_{\perp }^a+K_{\perp
}^b)+4K_{\parallel }]u^{{\bf j}}-
\nonumber    \\
&K&_{\perp}^a(u^{j_x+1,j_y}-u^{j_x-1,j_y})-K_{\perp }^b(u^{j_x,j_y+1}
-u^{j_x,j_y-1})=0.
\nonumber
\end{eqnarray}
This equation can be inverted to give: 
\begin{equation}
u^{{\bf j}}(x)=-\frac 1{4K_{\parallel }}
\sum_{{\bf j}^{\prime}}{\bf B}({\bf j}^{\prime }-{\bf j})
\sin \phi _0^{{\bf j}^{\prime }}(x) e^{-<\hat{\phi}^2>/2}
\label{uA3d}
\end{equation}
where:
\begin{equation}
{\bf B}({\bf j^\prime }-{\bf j})=\int_{-\pi }^\pi \int_{-\pi }^\pi 
\frac{d{\bf k}}{(2\pi )^2}\frac{\cos [{\bf k}\cdot ({\bf j^{\prime }-j})
]}{1+\epsilon^a \sin^2 \frac{k_x}2 +
\epsilon^b \sin^2 \frac{k_y}2}  
\label{Bint}
\end{equation}
where $\epsilon^{a,b}=\frac{K_{\perp }^{a,b}}{K_{\parallel }}$ 
are the relative interchain elastic couplings. 
The equations governing the classical fields $\phi _0^{{\bf j}}$ now 
become: 
\begin{eqnarray}
\partial_x^2\phi _0^{{\bf j}}&+&\frac 1{2\xi ^2}{\bf B}({\bf 0})\sin 
(2\phi_0^{{\bf j}})   \nonumber   \\
&+&\frac 1{\xi ^2}\sum_{{\bf j}^{^{\prime }}\neq 
{\bf j}}{\bf B}({\bf j}^{\prime }-{\bf j})\cos \phi _0^{{\bf j}}\sin 
\phi _0^{{\bf j}^{^{\prime}}}=0  
\label{soleq3d}
\end{eqnarray}
where we have only included for simplicity the effect of the NN
exchange interaction in each chain. In this situation, $\xi $
is given by Eq. (\ref{solwid}) with $g=0$. Moreover, as the
interchain elastic couplings do not affect the uniform dimerized 
state, the relation (\ref{width-gap}) between $\xi $ and the gap is 
preserved.

\begin{figure}
\psfig{figure=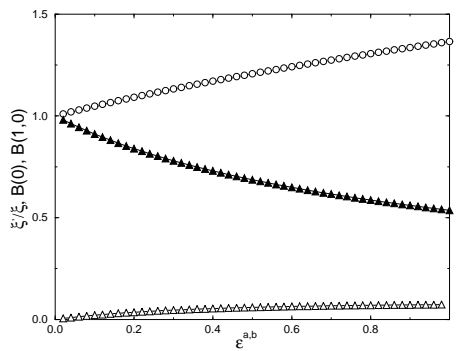,width=7.0cm,angle=-90}
\vspace{0.1cm}
\caption{The ratio between the effective soliton width and the
``bare" soliton width as a function of $\epsilon^a = \epsilon^b$ 
(circles).
Coefficients $B({\bf 0})$ (filled triangles) and $B(1,0)$
 (open triangles) given by Eq. (\ref{Bint}) are also shown.}
\label{fig2}
\end{figure}

The second term of Eq. (\ref{soleq3d}) is dominant due to the 
rapidly decaying effective interaction 
${\bf B}({\bf j^\prime - j})$
with the distance $|{\bf j}^{\prime }-{\bf j}|$. 
This behavior can be seen in Fig. \ref{fig2} for the case
$\epsilon^a = \epsilon^b$.
By neglecting the last term in Eq. (\ref{soleq3d})
we obtain a system of decoupled equations of 
the form (\ref{solsol}).
We consider a situation where a soliton is formed in a single chain 
in an otherwise SP state. Therefore we have 
$\phi_0^{{\bf j}}=\pi/2$ for $\bf j \ne \bf m$  and
$\sin (\phi_0^{{\bf j}})= \tanh (x/\xi^{\prime})$ for 
$\bf j = \bf m$.
The lattice pattern is obtained from Eq. (\ref{uA3d}): 

\begin{eqnarray}
u^{{\bf j}}= {\bf B(j- m)} u_0 \tanh (x/\xi^\prime)
+(1-{\bf B(j - m)}) u_0
\label{neighdist}
\end{eqnarray} 

This solution represents
a soliton at the origin of the {\bf m} chain.
It is easy to see that the displacement amplitude at one side of the
soliton is smaller than at the other side. This difference can be 
explained by the fact that at one side the dimerization has the
same phase with respect to that of the neighboring chains while at the
other these phases are opposite. At the same time, the neighbor chains 
$\bf j \ne \bf m$ feel the presence of the soliton in {\bf m}. 
As seen from Eq. (\ref{neighdist}) they are slightly distorted from the 
uniform dimerized pattern.

\vspace{-0.5cm}

From Eq. (\ref{soleq3d}) we obtain 
the effective soliton width $\xi^\prime=\xi / \sqrt{B({\bf 0})}$
which is {\it larger} than the ``bare" width $\xi $ of the one
chain problem previously discussed since $B({\bf 0}) < 1$.
We show in Fig. \ref{fig2} the behavior of 
$\xi^\prime/\xi$ as a function of $\epsilon^a$ for the case
of $\epsilon^a = \epsilon^b$.
It is difficult to extract from experimental data an estimation
of the values of $\epsilon^{a,b}$ since these are {\it effective}
elastic constants as introduced in the interaction term 
(\ref{helprime}).

After this work has been completed we became aware of a paper of 
Zang {\it et al.} \cite{Zang2} where the effects of frustration and
interchain magnetic coupling on the soliton width
were studied. Their conclusion about the frustration effects is
similar to the one presented here. Besides, 
they conclude that the large soliton width experimentally found in
CuGeO$_3$ is mainly due to the antiferromagnetic
interchain coupling. 
However, as the study of Ref. [\onlinecite{Zang2}]
is based on a mean-field approximation for the interchain coupling
and since
the soliton formation is a strongly local phenomena, we think
that the enhancement of the soliton width found by these authors
might be overestimated. In this case,
a contribution from the interchain
elastic coupling could be needed to explain the experimental result. 
 The question of to what extent these two mechanisms 
contribute to the soliton width could only be answered by studying
a model which includes both magnetic and elastic interchain
couplings.

\end{document}